%% file: main.tex
\documentclass[10pt,journal]{IEEEtran}

\input{setup/preamble.tex}

\input{setup/info.tex}

\begin{document}

\maketitle
\input{1_abstract.tex}

\input{2_introduction.tex}
\input{3_system_model.tex}
\input{7_results.tex}

\input{8_conclusion.tex}


\bibliographystyle{IEEEtran}
\bibliography{reference/mybib}

\end{document}

%% file: setup/preamble.tex
\usepackage{amssymb}
\usepackage{amsmath}
\usepackage{cite}
\usepackage{url}
\usepackage{xcolor}
\usepackage{cite,graphicx,amsmath,amssymb}
\usepackage{fancyhdr}
\usepackage{mdwmath}
\usepackage{mdwtab}
\usepackage{caption}
\usepackage{amsthm}
\usepackage{setspace}
\usepackage{hyperref}
\usepackage{algorithm}
\usepackage{algorithmic}
\usepackage{multirow}
\usepackage{makecell}
\usepackage{mathtools}
\usepackage{subcaption}
\usepackage{bm}
\usepackage{tikz}
\usepackage{xurl}

\hypersetup{colorlinks=true,
linkcolor=blue,
citecolor=blue,      
urlcolor=black,
}

\graphicspath{{./src/img/}}

\newtheorem{theorem}{Theorem}

\newtheorem{lemma}{Lemma}

\newtheorem{corollary}{Corollary}

\allowdisplaybreaks
\NewDocumentCommand{\multiubrace}{mmm}
 {
  \egreg_multiubrace:nnn {#1} {#2} {#3}
 }

\expandafter\def\expandafter\normalsize\expandafter{%
    \normalsize%
    \setlength\abovedisplayskip{4.5pt}%
    \setlength\belowdisplayskip{4.5pt}%
    \setlength\abovedisplayshortskip{2pt}%
    \setlength\belowdisplayshortskip{2pt}%
}

%% file: setup/info.tex
\title{\huge Wireless Sensing via Pinching-Antenna Systems}

\author{
        Zhaolin Wang,~\IEEEmembership{Member,~IEEE,}
        Chongjun Ouyang,~\IEEEmembership{Member,~IEEE,} 
        Yuanwei Liu,~\IEEEmembership{Fellow,~IEEE}, \\
        and Arumugam Nallanathan,~\IEEEmembership{Fellow,~IEEE}.

\thanks{Zhaolin Wang, Chongjun Ouyang, and Arumugam Nallanathan are with the School of Electronic Engineering and Computer Science, Queen Mary University of London, London E1 4NS, U.K. (e-mail: \{zhaolin.wang, c.ouyang, a.nallanathan\}@qmul.ac.uk). Yuanwei Liu is with the Department of Electrical and Electronic Engineering, The University of Hong Kong, Hong Kong (e-mail: yuanwei@hku.hk).}
\vspace{-0.7cm}
}

%% file: 1_abstract.tex
\begin{abstract}
    A wireless sensing architecture via pinching antenna systems is proposed. Compared to conventional wireless systems, PASS offers flexible antenna deployment and improved probing performance for wireless sensing by leveraging dielectric waveguides and pinching antennas (PAs). To enhance signal reception, leaky coaxial (LCX) cables are used to uniformly collect echo signals over a wide area. The Cramér-Rao bound (CRB) for multi-target sensing is derived and then minimized through the joint optimization of the transmit waveform and the positions of PAs. To solve the resulting highly coupled, non-convex problem, a two-stage particle swarm optimization (PSO)-based algorithm is proposed. Numerical results demonstrate significant gains in sensing accuracy and robustness over conventional sensing systems, highlighting the benefits of integrating LCX-based reception with optimized PASS configurations.
\end{abstract}

\begin{IEEEkeywords}
    Pinching-antenna system, waveform optimization, wireless sensing
\end{IEEEkeywords}

%% file: 2_introduction.tex
\section{Introduction} \label{sec:intro}

\IEEEPARstart{T}HE pinching antenna system (PASS) has recently attracted significant attention \cite{suzuki2022pinching, 10945421,wang2025modeling}. Unlike conventional wireless systems, PASS utilizes dielectric waveguides as the primary propagation medium and radiates signals into free space by pinching separate dielectric particles, referred to as pinching antennas (PAs), onto the waveguides. This design not only establishes strong line-of-sight channels and mitigates blockage issues but also enables high antenna reconfigurability through the flexible deployment of PAs. While this antenna deployment flexibility conceptually similar to fluid antennas \cite{9770295}, PASS achieves a much broader deployment range due to its use of dielectric waveguides. 

Given both the promising theoretical results \cite{10945421, wang2025modeling} and experimental evidence \cite{suzuki2022pinching} demonstrating the potential of PASS to enhance wireless communication performance, researchers have begun exploring its benefits for wireless sensing applications \cite{ding2025pinching,bozanis2025cram, khalili2025pinching}. In particular, the Cramér-Rao bound (CRB) of positioning error in a PASS-aided uplink system was analyzed in \cite{ding2025pinching}, revealing that PASS can significantly improve sensing accuracy. Building on this, \cite{bozanis2025cram} further confirmed the effectiveness of PASS in reducing the CRB in a bistatic wireless sensing scenario. Additionally, \cite{khalili2025pinching} proposed leveraging the large deployment range of PAs to provide enhanced target diversity, thereby further improving sensing performance.

Although several initial studies have explored the use of PASS for wireless sensing, key challenges remain unresolved. On one hand, some works propose utilizing PAs for signal reception \cite{ding2025pinching}, which potentially introduces two issues. First, while PAs offer strong advantages for transmission by focusing radiation energy on desired regions, their ability to capture a broad range of echo signals is limited due to sparse reception points. Second, deploying multiple PAs on a single waveguide for reception may lead to inter-antenna coupling effects according to coupled-mode theory \cite{wang2025modeling}, thereby increasing system complexity. On the other hand, alternative approaches employ conventional antenna arrays for signal reception \cite{bozanis2025cram, khalili2025pinching}, but these still suffer from high free-space pathloss and susceptibility to blockages. Furthermore, the joint optimization of transmit waveforms and PA placement to maximize multi-target sensing performance remains an open problem.

To address the aforementioned challenges, this letter proposes to use leaky coaxial (LCX) cables for signal reception in PASS-aided wireless sensing systems. Unlike dielectric waveguides, which use flexible PAs to concentrate probing power on specific target locations, LCX cables utilize densely and periodically spaced slots to efficiently capture reflected echo signals \cite{10871852}. Specifically, while PAs are well-suited for transmission by focusing energy toward intended targets, LCX cables are inherently advantageous for reception, as they can collect echo signals uniformly across a wider area. This complementary combination enables a win-win solution that enhances both transmission and reception performance in PASS-aided sensing. Furthermore, we investigate the joint optimization of transmit waveforms and PA positions to maximize multi-target sensing performance. To tackle the resulting highly coupled, non-convex problem, we proposed a two-stage particle swarm optimization (PSO)-based algorithm. Finally, numerical results validate the effectiveness of the proposed method and demonstrate the significant sensing performance gains of PASS over conventional wireless systems.

%% file: 3_system_model.tex
\vspace{-0.3cm}
\section{System Model} \label{sec:model}

\begin{figure}[t!]
    \centering
    \includegraphics[width=0.45\textwidth]{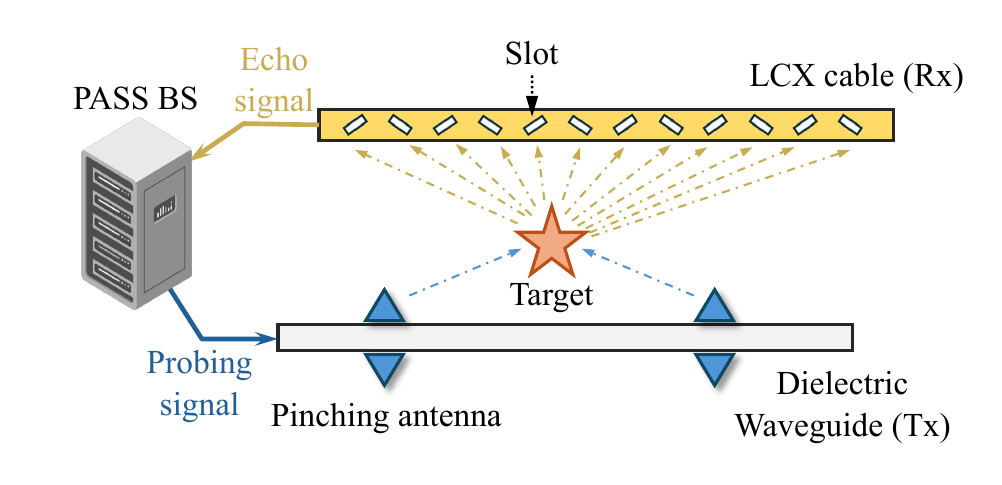}
    \caption{Illustration of the PASS-aided wireless sensing system.}
    \label{fig_system_model}
\end{figure}

As illustrated in Fig. \ref{fig_system_model}, we consider a narrowband PASS-aided wireless sensing system. In this system, $N$ dielectric waveguides are uniformly deployed throughout the region of interest, each connected to a radio frequency (RF) chain and equipped with $M_{\mathrm{t}}$ pinching antennas radiating probing signals to localize $K$ targets within the area. The probing signals reflected by the targets are collected at the base station via $N$ LCX cables, each positioned alongside its corresponding transmitting dielectric waveguide.  The lengths of both the dielectric waveguides and the LCX cables are denoted by $L$.


We assume the dielectric waveguides are deployed parallel to the $x$-axis. Thus, the position of the $m$-th PA on the $n$-th waveguide is represented by $\mathbf{p}_{n,m} = [x_{n,m}, y_{\mathrm{t},n}, z_{\mathrm{t,n}}]^T \in \mathbb{R}^{3 \times 1}$, where the value $x_{n,m}$ is adjustable by repositioning the PA along the waveguide. Targets are assumed to be located on the ground, with the $k$-th target's coordinate given by $\boldsymbol{\theta}_k = [\theta_{\mathrm{x},k}, \theta_{\mathrm{y}, k}, 0]^T \in \mathbb{R}^{3 \times 1}$. The distance between the $k$-th target and the $m$-th PA on the $n$-th waveguide is calculated as
\begin{equation}
    r_{k,n,m} = \sqrt{ (x_{n,m} - \theta_{\mathrm{x},k})^2 + (y_{\mathrm{t}, n} - \theta_{\mathrm{y},k})^2 + z_{\mathrm{t},n}^2 }.
\end{equation}        

Let $s_n(t) \in \mathbb{C}$ denote the signal input to the $n$-th waveguide. The signal impinging upon the $k$-th target from the $n$-th waveguide can thus be expressed as \cite{wang2025modeling}
\begin{equation}
    \widetilde{\varphi}_{n,k}(t) = \widetilde{\mathbf{a}}^T(\boldsymbol{\theta}_k, \mathbf{x}_n) \mathbf{g}(\mathbf{x}_n) s_n(t),
\end{equation}  
where $\mathbf{x}_n = [x_{n,1},\cdots,x_{n,M_{\mathrm{t}}}]^T \in \mathbb{R}^{M_{\mathrm{t}} \times 1}$ represents the position vector of the PAs on the $n$-th waveguide, and $\widetilde{\mathbf{a}}(\boldsymbol{\theta}_k, \mathbf{x}_n) \in \mathbb{C}^{M_{\mathrm{t}} \times 1}$ and $\mathbf{g}(\mathbf{x}_n) \in \mathbb{C}^{M_{\mathrm{t}} \times 1}$ denote the \emph{free-space} and \emph{in-waveguide} propagation vectors, respectively. More specifically, the free-space propagation vector is given by 
\begin{equation}
    \widetilde{\mathbf{a}}(\boldsymbol{\theta}_k, \mathbf{x}_n) = \left[ \frac{\eta e^{-j \frac{2\pi}{\lambda} r_{k,n,1}} }{r_{k,n,1}}, \cdots, \frac{\eta e^{-j \frac{2\pi}{\lambda} r_{k,n,M_{\mathrm{t}}}} }{r_{k,n,M_{\mathrm{t}}}} \right]^T,
\end{equation}
where $\eta = \frac{\lambda}{4 \pi}$ denotes pathloss factor. Furthermore, the in-waveguide propagation vector is given by 
\begin{equation}
    \mathbf { g } ( \mathbf { x } _ { n } ) = \left[ \alpha _ { 1 } e ^ { - j \frac{2 \pi n_{\mathrm{t}}}{\lambda} x _ { n, 1 } }, \ldots, \alpha _ { M_{\mathrm{t}} } e ^ { - j \frac{2 \pi n_{\mathrm{t}}}{\lambda} x _ { n, M_{\mathrm{t}} } } \right] ^ { T },
\end{equation}
where $\alpha_m \in \mathbb{R}$ determines the ratio of power radiated from the $m$-th PA, and $n_{\mathrm{t}} \in \mathbb{R}$ denotes the effective refractive index of the dielectric waveguide. Then, the signal impinged upon the $k$-th target from all waveguides can be expressed as \cite{wang2025modeling}
\begin{align}
    \varphi_k(t) = & \sum_{n=1}^N \widetilde{\mathbf{a}}^T(\boldsymbol{\theta}_k, \mathbf{x}_n) \mathbf{g}(\mathbf{x}_n) s_n(t), \nonumber \\
    = & \mathbf{a}^T(\boldsymbol{\theta}_k, \mathbf{X}) \mathbf{G}(\mathbf{X}) \mathbf{s}(t),
\end{align} 
where $\mathbf{X} = [\mathbf{x}_1,\cdots,\mathbf{x}_N]^T$, $\mathbf{s}(t) = [s_1(t),\cdots,s_N(t)]^T$, $\mathbf{G}(\mathbf{X}) = \mathrm{blkdiag}\left\{ \mathbf{g}(\mathbf{x}_1),\cdots, \mathbf{g}(\mathbf{x}_N)\right\}$, and $\mathbf{a}(\boldsymbol{\theta}_k, \mathbf{X}) = \left[ \widetilde{\mathbf{a}}^T(\boldsymbol{\theta}_k, \mathbf{x}_1), \cdots, \widetilde{\mathbf{a}}^T(\boldsymbol{\theta}_k, \mathbf{x}_{N_{\mathrm{t}}}) \right]^T$.  


The signal $\varphi_k(t)$ reflected by the target is subsequently received by the LCX cables for further processing. Let $d$ denote the spacing between the slots, implying totally $M_{\mathrm{r}} = L/d + 1$ slots on each LCX cable. Therefore, the position of the $m$-th slot on the $n$-th LCK cable can be denoted by $\widetilde{\mathbf{p}} = [ (m-1)d, y_{\mathrm{r},n}, z_{\mathrm{r},n} ]^T$. The following the modeling of the transmit signal, the overall echo signal from $k$-th target can be expressed as \cite{10871852}
\begin{equation}
    \widetilde{\mathbf{y}}_k(t) = \beta_k \mathbf{V}^T \mathbf{b}(\boldsymbol{\theta}_k) \varphi_k(t),
\end{equation} 
where $\beta_k \in \mathbb{C}$ denotes the reflection coefficient, and $\mathbf{V} \in \mathbb{C}^{N M_{\mathrm{r}} \times N}$ and $\mathbf{b}(\boldsymbol{\theta}_k)$ account for the \emph{free-space} and \emph{in-waveguide} propagation, respectively, defined as follows:  
\begin{align}
    \mathbf{V} = & \mathbf{I}_{N} \otimes \left[1,  e^{-j \frac{2 \pi n_{\mathrm{r}}}{\lambda} d},\cdots,e^{-j \frac{2 \pi n_{\mathrm{r}}}{\lambda} (M_{\mathrm{r}}-1) d} \right]^T, \\
    \mathbf{b}(\boldsymbol{\theta}_k) = & \left[\widetilde{\mathbf{b}}_1^T(\boldsymbol{\theta}_k),\cdots,\widetilde{\mathbf{b}}_{N}^T(\boldsymbol{\theta}_k) \right]^T,\\
    \widetilde{\mathbf{b}}_n(\boldsymbol{\theta}_k) = & \left[ \frac{\eta e^{-j \frac{2\pi}{\lambda} \widetilde{r}_{k,n,1} }}{\widetilde{r}_{k,n,1}}, \cdots, \frac{\eta e^{-j \frac{2\pi}{\lambda} \widetilde{r}_{k,n,M_{\mathrm{r}}} }}{\widetilde{r}_{k,n,M_{\mathrm{r}}}} \right]^T, \\
    \widetilde{r}_{k,n,m} = & \sqrt{((m-1)d  - \theta_{\mathrm{x},k})^2 + (y_{\mathrm{r},n} - \theta_{\mathrm{y},k})^2 + z_{\mathrm{r},n}^2}.
\end{align}
The overall echo signal received from $K$ targets is given by 
\begin{align}
    \mathbf{y}(t) = & \sum_{k=1}^K \beta_k \mathbf{V}^T \mathbf{b}(\boldsymbol{\theta}_k) \varphi_k(t) + \mathbf{z}(t) \nonumber \\ 
    = & \mathbf{V}^T \mathbf{B}(\boldsymbol{\theta}) \mathbf{\Phi} \mathbf{A}^T(\boldsymbol{\theta}, \mathbf{X}) \mathbf{G}(\mathbf{X}) \mathbf{s}(t) + \mathbf{z}(t),
\end{align}
where $\boldsymbol{\theta} = [\boldsymbol{\theta}_1^T,\cdots,\boldsymbol{\theta}_K^T]^T$, $\mathbf{A}(\boldsymbol{\theta}, \mathbf{X}) = [ \mathbf{a}(\boldsymbol{\theta}_1, \mathbf{X}), \cdots, $ $\mathbf{a}(\boldsymbol{\theta}_K, \mathbf{X}) ]$, $\mathbf{B}(\boldsymbol{\theta}) = [\mathbf{b}(\boldsymbol{\theta}_1),\cdots,\mathbf{b}(\boldsymbol{\theta}_K)]$, $\boldsymbol{\beta} = [\beta_1,\cdots,\beta_K]^T$, and $\mathbf{\Phi} = \mathrm{diag}(\boldsymbol{\beta})$. 
Furthermore, $\mathbf{z}(t) \sim \mathcal{CN}(0, \sigma^2 \mathbf{I}_N)$ denotes the complex Gaussian noise.

\section{Cramér-Rao Bound Minimization}

In this article, we aim to jointly design the transmit waveform $\mathbf{s}(t)$, or equivalently, its covariance matrix $\mathbf{R}_s = \frac{1}{T} \sum_{t=1}^T \mathbf{s}(t) \mathbf{s}^H(t)$ \cite{4359542}, and the position $\mathbf{X}$ of PAs to optimize the accuracy of target localization based on the received echo signal $ \mathbf{Y} = [\mathbf{y}(1),\cdots,\mathbf{y}(T)]$ over a coherent processing interval of length $T$. The CRB, which characterizes the lower bound on the estimation error achievable by unbiased estimators, is an effective performance metric for the sensing system design~\cite{4359542},~\cite{9652071}. 

\vspace{-0.3cm}
\subsection{Problem Formulation}
We first derive the CRB in the following. Note that the overall real-valued vector of the unknown target parameters to estimate from $\mathbf{Y}$ is $\boldsymbol{\xi} = [ \boldsymbol{\theta}^T, \Re(\boldsymbol{\beta}^T), \Im(\boldsymbol{\beta}^T) ]^T \in \mathbb{R}^{4K \times 1}$. Let $\mathbf{F} \in \mathbb{C}^{4K \times 4K}$ denote the Fisher information matrix (FIM) for the estimation of $\boldsymbol{\xi}$. To calculate the FIM, we first define the following matrices:
\begin{align}
    \mathbf{A}_d =  & \mathbf{A} \otimes [1,1], \mathbf{B}_d =  \mathbf{B} \otimes [1,1], \mathbf{\Phi}_d = \mathbf{\Phi} \otimes \mathbf{I}_2, \\ 
    \dot{\mathbf{A}}_d = & \left[ \cdots \frac{\partial \mathbf{a}(\boldsymbol{\theta}_k, \mathbf{X})}{\partial \theta_{\mathrm{x},k}}, \frac{\partial \mathbf{a}(\boldsymbol{\theta}_k, \mathbf{X})}{\partial \theta_{\mathrm{y},k}} \cdots\right], \\
    \dot{\mathbf{B}}_d = & \left[ \cdots \frac{\partial \mathbf{b}(\boldsymbol{\theta}_k)}{\partial \theta_{\mathrm{x}, k}}, \frac{\partial \mathbf{b}(\boldsymbol{\theta}_k)}{\partial \theta_{\mathrm{y}, k}} \cdots  \right].
\end{align}
The partial derivatives in the above expression can be calculated based on the following results for $i \in \{ \mathrm{x}, \mathrm{y} \}$ :
\begin{align}
    & \frac{\partial \mathbf{a}(\boldsymbol{\theta}_k, \mathbf{X})}{\partial \theta_{i,k}} = \left[ \frac{\partial \widetilde{\mathbf{a}}^T(\boldsymbol{\theta}_k, \mathbf{x}_1)}{\partial \theta_{i,k}},\cdots,\frac{\partial \widetilde{\mathbf{a}}^T(\boldsymbol{\theta}_k, \mathbf{x}_N)}{\partial \theta_{i,k}}  \right]^T, \\
    & \frac{\partial \mathbf{b}(\boldsymbol{\theta}_k)}{\partial \theta_{i,k}} = \left[ \frac{\partial \widetilde{\mathbf{b}}_1^T(\boldsymbol{\theta}_k)}{\partial \theta_{i,k}},\cdots,\frac{\partial \widetilde{\mathbf{b}}_N^T(\boldsymbol{\theta}_k)}{\partial \theta_{i,k}}  \right]^T,  \\
    & \left[\frac{\partial \widetilde{\mathbf{a}}(\boldsymbol{\theta}_k, \mathbf{x}_n)}{\partial \theta_{i,k}}\right]_m  = \frac{ \eta \kappa_{i,n,m} \left(j \frac{2\pi}{\lambda} + 1\right) e^{-j \frac{2\pi}{\lambda} r_{k,n,m}}  }{r_{k,n,m}^3}, \\
    & \left[\frac{\partial \widetilde{\mathbf{b}}_n(\boldsymbol{\theta}_k)}{\partial \theta_{i,k}}\right]_m  = \frac{ \eta \widetilde{\kappa}_{i,n,m} \left(j \frac{2\pi}{\lambda} + 1\right) e^{-j \frac{2\pi}{\lambda} \widetilde{r}_{k,n,m}}  }{\widetilde{r}_{k,n,m}^3},
\end{align}
where $\kappa_{\mathrm{x},n,m} = (x_{n,m} - \theta_{\mathrm{x},k})$, $\kappa_{\mathrm{y},n,m} = (y_{\mathrm{t},m} - \theta_{\mathrm{y},k})$, $\widetilde{\kappa}_{\mathrm{x},n,m} = ((m-1)d - \theta_{\mathrm{x},k})$, and $\kappa_{\mathrm{y}n,m} = (y_{\mathrm{r},m} - \theta_{\mathrm{y},k})$.
Subsequently, the FIM can be calculated following \cite[Appendix A]{4359542} as    
\begin{equation}
    \mathbf{F} = \frac{2 T}{\sigma^2} \begin{bmatrix}
        \Re(\mathbf{F}_{11}) & \Re(\mathbf{F}_{12}) & -\Im(\mathbf{F}_{12}) \\
        \Re(\mathbf{F}_{12}^T) & \Re(\mathbf{F}_{22}) & -\Im(\mathbf{F}_{22}) \\
        - \Im(\mathbf{F}_{12}^T) & -\Im(\mathbf{F}_{22}^T) & \Re(\mathbf{F}_{22})
    \end{bmatrix}.
\end{equation} 
The matrices $\mathbf{F}_{11} \in \mathbb{C}^{2K \times 2K}$, $\mathbf{F}_{12} \in \mathbb{C}^{2K \times K}$, and $\mathbf{F}_{22} \in \mathbb{C}^{K \times K}$ are given by
\begin{align}
    \mathbf{F}_{11} & = \left(\dot{\mathbf{B}}_d^H \mathbf{V}^* \mathbf{V}^T \dot{\mathbf{B}}_d  \right) \odot \left( \mathbf{\Phi}_d^* \mathbf{A}_d^H \mathbf{G}^* \mathbf{R}_s^* \mathbf{G}^T \mathbf{A}_d \mathbf{\Phi}_d \right) \nonumber \\
    & +  \left(\dot{\mathbf{B}}_d^H \mathbf{V}^* \mathbf{V}^T \mathbf{B}_d  \right) \odot \left( \mathbf{\Phi}_d^* \mathbf{A}_d^H \mathbf{G}^* \mathbf{R}_s^* \mathbf{G}^T \dot{\mathbf{A}}_d \mathbf{\Phi}_d \right) \nonumber \\
    & + \left( \mathbf{B}_d^H \mathbf{V}^* \mathbf{V}^T \dot{\mathbf{B}}_d  \right) \odot \left( \mathbf{\Phi}_d^* \dot{\mathbf{A}}_d^H \mathbf{G}^* \mathbf{R}_s^* \mathbf{G}^T \mathbf{A}_d \mathbf{\Phi}_d \right) \nonumber \\
    & + \left( \mathbf{B}_d^H \mathbf{V}^* \mathbf{V}^T \mathbf{B}_d  \right) \odot \left( \mathbf{\Phi}_d^* \dot{\mathbf{A}}_d^H \mathbf{G}^* \mathbf{R}_s^* \mathbf{G}^T \dot{\mathbf{A}}_d \mathbf{\Phi}_d \right), \\
    \mathbf{F}_{12} & =\left(\dot{\mathbf{B}}_d^H \mathbf{V}^* \mathbf{V}^T \mathbf{B}  \right) \odot \left( \mathbf{\Phi}_d^* \mathbf{A}_d^H \mathbf{G}^* \mathbf{R}_s^* \mathbf{G}^T \mathbf{A} \right) \nonumber \\
    & + \left(\mathbf{B}_d^H \mathbf{V}^* \mathbf{V}^T \mathbf{B}  \right) \odot \left( \mathbf{\Phi}_d^* \dot{\mathbf{A}}_d^H \mathbf{G}^* \mathbf{R}_s^* \mathbf{G}^T\mathbf{A} \right), \\
    \mathbf{F}_{22} & = \left(\mathbf{B}^H \mathbf{V}^* \mathbf{V}^T \mathbf{B}  \right) \odot \left( \mathbf{A}^H \mathbf{G}^* \mathbf{R}_s^* \mathbf{G}^T \mathbf{A} \right).
\end{align}

The CRBs for the estimation of $\boldsymbol{\theta}$ correspond to the first $2K$ diagonal entries of $\mathbf{F}^{-1}$. Therefore, using the block matrix inversion lemma, the CRBs for the estimation of $\boldsymbol{\theta}$ is also corresponding to the all diagonal entries of the following CRB matrix:
\begin{equation}
    \mathrm{CRB}(\boldsymbol{\theta}) = \left( \Re(\mathbf{F}_{11}) - \widetilde{\mathbf{F}}_{12} \widetilde{\mathbf{F}}_{22}^{-1} \widetilde{\mathbf{F}}_{12}^T \right)^{-1},
\end{equation}
where 
\begin{align}
    \widetilde{\mathbf{F}}_{12} = & \left[ \Re(\mathbf{F}_{12}), -\Im(\mathbf{F}_{12})  \right], \\
    \widetilde{\mathbf{F}}_{22} = & \begin{bmatrix}
        \Re(\mathbf{F}_{22}) & -\Im(\mathbf{F}_{22}) \\
        -\Im(\mathbf{F}_{22}^T) & \Re(\mathbf{F}_{22})
    \end{bmatrix}.
\end{align}
Therefore, the corresponding CRB minimization problem can be formulated as \cite{4359542, wang2025modeling}
\begin{subequations} 
    \label{opt_problem}
    \begin{align}
        \min_{\mathbf{X}, \mathbf{R}_s} \quad & \mathrm{Tr}(\mathrm{CRB}(\boldsymbol{\theta})) \\
        \mathrm{s.t.} \quad & \mathrm{Tr}(\mathbf{R}_s) = P_{\mathrm{t}}, \mathbf{R}_s \succeq 0, \\
        & x_{n,m} - x_{n,m-1} \ge \Delta x, \forall n, m \neq 1, \\
        & 0 \le x_{n,m} \le L, \forall n, m,
    \end{align}
\end{subequations}
where $P_{\mathrm{t}}$ denotes the transmit power and $\Delta x$ denotes the minimum spacing between two adjacent PAs on the same waveguide.

\subsection{Proposed Solution}
The optimization problem \eqref{opt_problem} is highly coupled and non-convex. In the following, we propose a two-stage optimization method to solve it. 

\subsubsection{Optimizing Position of PAs}

In the first stage, the positions $\mathbf{X}$ of the PAs are optimized. To maximize the effectiveness of this optimization, the transmit covariance matrix is set as $\mathbf{R}_s = \frac{P}{N} \mathbf{I}_N$.
This choice represents an unbiased waveform design, ensuring that the system relies solely on pinching beamforming to enhance the sensing performance, and thus guaranteeing the positions of the PAs are effectively and sufficiently optimized. However, solving problem \eqref{opt_problem} with respect to $\mathbf{X}$ remains challenging due to the ill-conditioned and highly multimodal nature of the objective function. This complexity arises from waveguide sharing and the large-scale deployment range of the PAs, which together result in numerous local optima. 

PSO is an efficient method for solving such problems, offering a low-complexity global search capability \cite{985692}. To apply PSO, we reformulate problem \eqref{opt_problem} with respect to $\mathbf{X}$ into the following unconstrained equivalent form:
\begin{equation} \label{opt_problem_new}
    \min_{\mathbf{X}} \quad \mathrm{Tr}(\mathrm{CRB}(\boldsymbol{\theta})) + \mathbb{I}(\mathbf{X}),
\end{equation}
where $\mathbb{I}(\mathbf{X})$ is a feasibility indicator function defined as
\begin{equation}
    \mathbb{I}(\mathbf{X}) = \begin{cases}
        0, & \text{if } \mathbf{X} \in \mathcal{S}, \\
        + \infty, & \text{otherwise},
    \end{cases}
\end{equation}
and $\mathcal{S}$ denotes the feasible region:
\begin{equation}
    \mathcal{S} = \left\{  \mathbf{X} \hspace{0.1cm} \Bigg| \begin{matrix}
        & x_{n,m} - x_{n,m-1} \ge \Delta x, \forall n, m \neq 1, \\
        & 0 \le x_{n,m} \le L, \forall n, m.
    \end{matrix} \right\}.
\end{equation} 
Although problem \eqref{opt_problem_new} is non-differentiable, PSO can still be applied efficiently. As a widely used and mature global optimization technique, PSO is readily available through standard toolboxes, such as the Global Optimization Toolbox in MATLAB. Thus, the implementation details are omitted here.

\subsubsection{Optimizing Transmit Waveform}

In the second stage, the transmit waveform covariance matrix $\mathbf{R}_s$ is further optimized based on the previously determined PA positions. Although the original problem \eqref{opt_problem} with respect to $\mathbf{R}_s$ is non-convex, it can be equivalently reformulated into a convex optimization problem, as shown in \cite[Proposition 1]{10050406}:
\begin{subequations} 
    \begin{align}
        \min_{\mathbf{U}, \mathbf{R}_s} \quad & \mathrm{Tr} \left(\mathbf{U}^{-1}\right) \\
        \mathrm{s.t.} \quad & \begin{bmatrix}
            \Re(\mathbf{F}_{11}) - \mathbf{U} & \widetilde{\mathbf{F}}_{12} \\
            \widetilde{\mathbf{F}}_{12}^T & \widetilde{\mathbf{F}}_{22}
        \end{bmatrix} \succeq 0, \\
        & \mathrm{Tr}(\mathbf{R}_s) = \mathrm{P}_{\mathrm{t}}, \\
        & \mathbf{R}_s \succeq 0, \mathbf{U} \succeq 0,
    \end{align}
\end{subequations}
where $\mathbf{U} \in \mathbb{C}^{2K \times 2K}$ is an auxiliary matrix. This reformulated problem is convex, as the objective function—trace of the inverse of a positive semidefinite matrix—is convex, and all constraints are affine in the optimization variables. Therefore, this problem can be solved optimally by standard convex program solvers such as CVX \cite{cvx}. 

%% file: 7_results.tex
\section{Numerical Results} \label{sec:results}
 
\begin{figure*}[t!]
  \centering
  \begin{minipage}[t]{0.33\textwidth}
    \centering
    \includegraphics[width=1\textwidth]{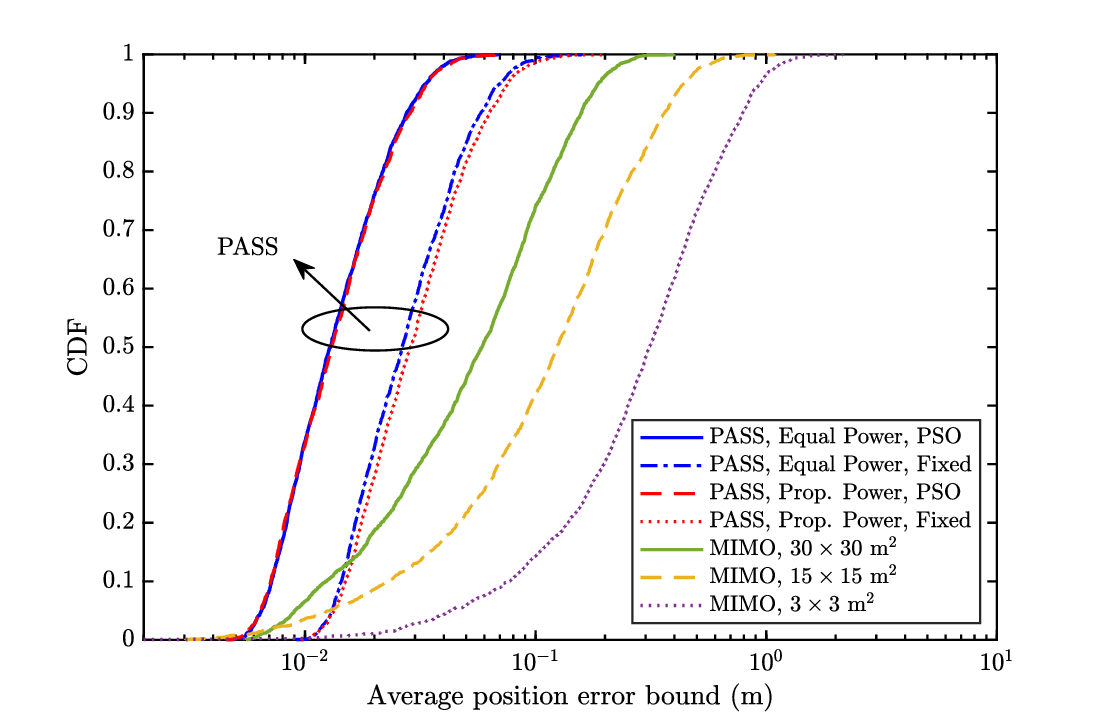}
    \caption{Cumulative distribution function of the position error bound.}
    \label{fig_CDF}
  \end{minipage}
  \begin{minipage}[t]{0.66\textwidth}
    \begin{subfigure}[t]{0.5\textwidth}
      \includegraphics[width=1\textwidth]{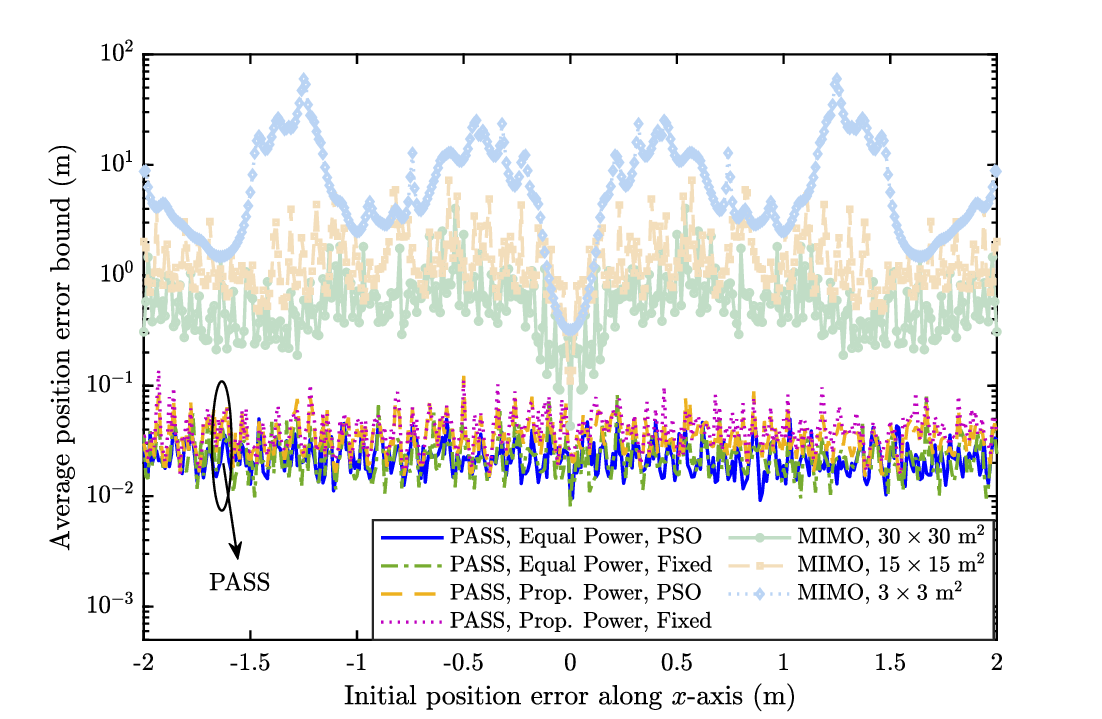}
      \caption{$x$-axis error.}
      \label{fig_uncertentity_x}
    \end{subfigure}
    \begin{subfigure}[t]{0.5\textwidth}
      \includegraphics[width=1\textwidth]{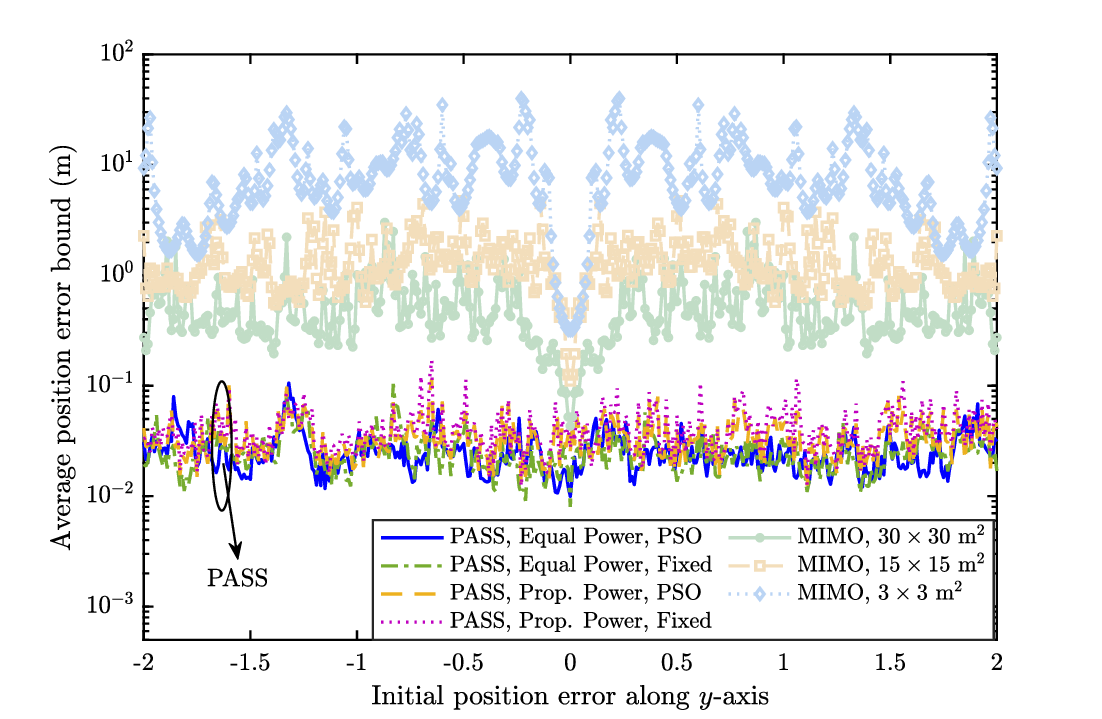}
      \caption{$y$-axis error.}
      \label{fig_uncertentity_y}
    \end{subfigure}
    \caption{Position error bound versus the initial estimation position error.}
    \label{fig_uncertentity}
  \end{minipage}
  \vspace{-0.3cm}
\end{figure*}

In this section, numerical results are presented to demonstrate the advantages of PASS for wireless sensing and to validate the effectiveness of the proposed algorithms. Unless otherwise specified, the following simulation setup is used. We consider a region of interest defined by $x \in [0, 30]$ m and $y \in [0, 30]$ m with $K = 2$ targets. Within this region, $N = 5$ dielectric waveguides and LCX cables are deployed for signal transmission and reception, respectively, each with a maximum length of $L = 30$ m and height $z_{\mathrm{t},n} = z_{\mathrm{r},n} = 3$ m. The $y$-coordinates of the $n$-th dielectric waveguide and LCX cable are set to $y_{\mathrm{t},n} = (5n - 0.5)$ m and $y_{\mathrm{r},n} = (5n + 0.5)$ m, respectively. Their effective refractive indexes are set to $n_{\mathrm{t}} = 1.4$ and $n_{\mathrm{r}} = 1.1$ \cite{9084280, 10945421}. Each waveguide is equipped with $M_{\mathrm{t}} = 4$ PAs with a minimum spacing of $\Delta x = 0.3$ m, and the spacing between adjacent slots on the LCX cables is set to $d = 0.08$ m \cite{10871852}. 

The carrier frequency, transmit power, noise power, and number of snapshots are set to $f = 15$ GHz, $P_{\mathrm{t}} = 20$ dBm, $\sigma^2 = -80$ dBm, and $T = 256$, respectively. Two PA power models are considered: the \emph{equal-power} model and the \emph{proportional-power} model, both satisfying the constraint $\sum_{i=1}^{M_{\mathrm{t}}} \alpha_m^2 = 0.9$. We refer to \cite{wang2025modeling} for the details of these two power models due to space limitation. Finally, the number of particles in the PSO algorithm is set to $30$.

For performance comparison, we consider following two benchmarks. 1) \textbf{Fixed PAs}: In this benchmark, the PAs are fixed and uniformly deployed along each waveguide, with positions given by $x_{n,m} = \frac{(m-1)L}{M-1}, \forall n, m$. 2) \textbf{Conventional MIMO}: This scheme assumes a fully digital uniform planar array (UPA) consisting of $10 \times 10$ antennas, used to transmit probing signals and receive echo signals. The array is parallel to the $x\textendash y$ plane at a height of $z = 3$ m, and centered at the point $(15, 15, 0)$ m.  In addition, we consider different aperture sizes for UPAs in conventional MIMO systems, where $a \times a$ indicates that the UPA spans a length of $a$ along both the $x$- and $y$-axes.    

Fig. \ref{fig_CDF} presents the cumulative distribution function (CDF) of the average position error bound (PEB). The average PEB is defined as $\mathrm{PEB} = \frac{1}{K} \sum_{k=1}^K \sqrt{ \mathrm{CRB}_{\theta_{\mathrm{x},k}} + \mathrm{CRB}_{\theta_{\mathrm{y},k}}}$,
where $\mathrm{CRB}_{\theta_{\mathrm{x},k}}$ and $\mathrm{CRB}_{\theta_{\mathrm{y},k}}$ are CRBs for estimating $\theta_{\mathrm{x},k}$ and $\theta_{\mathrm{y},k}$, respectively. The CDF is obtained by simulating over $2000$ random samples. It can be observed that, compared to conventional MIMO, the proposed PASS system significantly enhances both the average performance and the stability of wireless sensing. This improvement is attributed to the use of dielectric waveguides and LCX cables, which reduce path loss and extend coverage. Moreover, optimizing the positions of PAs using PSO further improves sensing performance, underscoring the importance of leveraging PA reconfigurability to concentrate radiation energy on target locations. Finally, for conventional MIMO systems, increasing the aperture size enhances sensing performance due to stronger near-field effects, which improve localization capability.

The optimization criteria used in this paper are based on the CRB matrix, which in practice need to be estimated using initial parameter estimates. Therefore, following \cite{9652071}, we evaluate the impact of initial position estimation errors in Fig. \ref{fig_uncertentity}, where we consider two targets located at $(5, 7.5)$ m and $(25, 12.5)$ m, respectively. It can be observed that PASS-aided wireless sensing, regardless of whether the PA positions are optimized or fixed, demonstrates significant robustness to initial estimation errors due to reduced path loss and enhanced coverage. In contrast, conventional MIMO systems exhibit noticeable performance degradation as the initial position estimation error increases.



%% file: 8_conclusion.tex
\vspace{-0.3cm}
\section{Conclusions} \label{sec:conclusion}

This letter has proposed a PASS-aided wireless sensing system with LCX-based signal reception and joint waveform and PA position optimization. The proposed approach significantly enhances sensing accuracy and robustness compared to conventional methods.